\def\tr{{\rm tr}\,}

\def\b{\bibitem}
\def\be{\begin{equation}}
\def\ee{\end{equation}}
\def\bea{\begin{eqnarray}}
\def\eea{\end{eqnarray}}
\def\bml{\begin{mathletters}}
\def\eml{\end{mathletters}}
\documentstyle[aps,prb,eqsecnum,psfig,epsf,floats]{revtex}
\draft
\begin{document}
\def\SNG{{\em Physical Review Style and Notation Guide}}
\def\LUG {{\em \LaTeX{} User's Guide \& Reference Manual}}
\def\btt#1{{\tt$\backslash$\string#1}}%
\def\REVTeX{REV\TeX}
\def\AmS{{\protect\the\textfont2
        A\kern-.1667em\lower.5ex\hbox{M}\kern-.125emS}}
\def\AmSLaTeX{\AmS-\LaTeX}
\def\BibTeX{\rm B{\sc ib}\TeX}
\twocolumn[\hsize\textwidth\columnwidth\hsize\csname@twocolumnfalse%
\endcsname
\title{Quantum critical behavior in disordered itinerant ferromagnets:\\
       Logarithmic corrections to scaling\\
       \small{$[$ Phys. Rev. B {\bf 63}, 174428 (2001) $]$}
                }
\author{D.Belitz}
\address{Department of Physics and Materials Science Institute\\
         University of Oregon,
         Eugene, OR 97403}
\author{T.R.Kirkpatrick}
\address{Institute for Physical Science and Technology, and Department of
         Physics\\
         University of Maryland,
         College Park, MD 20742}
\author{Maria Teresa Mercaldo}
\address{Institute for Physical Science and Technology, and Department of
         Physics\\
         University of Maryland,
         College Park, MD 20742\\
         and\\
         Dipartimento di Scienze Fisiche ``E.R. Caianiello'' and Istituto
         Nazionale di Fisica per la Materia,\\
         Universit{\'a} di Salerno, I-84081 Baronissi (SA), Italy}
\author{Sharon L. Sessions}
\address{Department of Physics and Materials Science Institute\\
         University of Oregon,
         Eugene, OR 97403}
\date{\today}
\maketitle

\begin{abstract}
The quantum critical behavior of disordered itinerant ferromagnets is
determined exactly by solving a recently developed effective field theory. 
It is shown that there are logarithmic corrections to a previous calculation
of the critical behavior, and that the exact critical behavior coincides
with that found earlier for a phase transition of undetermined nature in
disordered interacting electron systems. This confirms a previous suggestion 
that the unspecified transition should be identified with the ferromagnetic 
transition. The behavior of the conductivity, the tunneling density of states, 
and the phase and quasiparticle relaxation rates
across the ferromagnetic transition is also calculated.
\end{abstract}
]
\section{Introduction}
\label{sec:I}

In a recent paper,\cite{us_paper_I} hereafter denoted by I, a local field 
theory capable of describing the zero temperature ($T=0$) ferromagnetic phase 
transition in disordered itinerant electron systems was developed. In the 
present paper this theory is used to exactly determine the critical behavior 
at the phase transition, and the connections between the local theory and 
previous descriptions of the ferromagnetic quantum phase transition are 
established.

Historically, the ferromagnetic transition in itinerant electron systems
at $T=0$ was the first quantum phase transition to be
studied in detail. Hertz\cite{Hertz} concluded that the transition in the 
physically 
interesting dimension $d=3$ was mean field-like. The basic idea behind this 
result was that the effective dimension of the system, which is given by 
the spatial dimension $d$ plus the effective time dimension $z$, was above the
upper critical dimension for the transition, so that fluctuation effects 
could be ignored. This conclusion is now known to be incorrect. For example, 
the Harris criterion\cite{Harris} for phase transitions in disordered systems 
states that the correlation length exponent $\nu$ must satisfy the inequality 
$\nu \geq 2/d$, whereas mean-field theory gives $\nu =1/2$ for all $d$. 
This in turn implies that a simple mean-field description 
{\em must} break down in dimensions $d<4$.

The reason for this breakdown of mean-field theory was shown in 
Ref.\ \onlinecite{us_dirty} to be the existence, in itinerant electron 
systems, of soft or massless modes other than the order parameter fluctuations,
which were not taken into account in Hertz's theory. In disordered systems 
these modes are diffusive, and they couple to the order parameter
fluctuations and modify the critical behavior.\cite{clean_footnote}
Among other things, they lead to a correlation length exponent that satisfies 
the Harris criterion. Technically, if these additional soft modes are 
integrated out, they lead to
a long-ranged interaction between the order parameter fluctuations. It was
argued that once this effect is taken into account, all other fluctuation
effects are suppressed by the long-range nature of the interactions and that
the critical behavior is governed by a fixed point that is Gaussian, but
does not yield mean-field exponents. It was thought that the critical
behavior found in Ref.\ \onlinecite{us_dirty} was exact.

Several years before the work reported in Ref.\ \onlinecite{us_dirty},
the study of metal-insulator transitions of disordered interacting 
electrons constituted a separate development in the many-electron 
problem.\cite{us_R} Within this context, a transition was encountered 
that was {\it not} a metal-insulator transition, but rather of magnetic 
nature.\cite{us_IFS} Due to the methods used in Ref.\ \onlinecite{us_IFS}, 
the order parameter and the nature
of the ordered state were not identified, but the critical behavior
was determined and found to consist of power laws with simple exponents,
modified by complicated logarithmic corrections. The critical behavior
for the ferromagnetic transition determined in Ref.\ \onlinecite{us_dirty}
turned out to consist of the same simple power laws, albeit with different,
and much simpler, logarithmic corrections. This led, in 
Ref.\ \onlinecite{us_dirty}, to the suggestion that the transition
studied in Ref.\ \onlinecite{us_IFS} was in fact the ferromagnetic 
transition. The discrepancy in the logarithmic corrections between the
two approaches was attributed to the fact that of the two integral
equations derived in Ref.\ \onlinecite{us_IFS}, only one had been shown
to be exact. The conclusion thus was that the theories presented in 
Refs.\ \onlinecite{us_dirty} and \onlinecite{us_IFS} had treated the same
problem, and that the former solution was exact, while the latter was
approximate.

The latter conclusion, however, relied on a weak link in the chain of
arguments, since the theory developed in Ref.\ \onlinecite{us_dirty}
was not very suitable for determining logarithmic corrections to
power laws. The reason was that the additional soft modes
were integrated out to obtain a description solely in terms of order 
parameter fluctuations. The resulting field theory was thus nonlocal, which
makes explicit calculations cumbersome. Consequently, most of the
arguments used in Ref.\ \onlinecite{us_dirty} to determine the critical
behavior were simple power counting techniques that were not sensitive
to logarithmic corrections to power laws. 

It is the purpose of the present
paper, in conjunction with the preceding paper I, to settle the remaining
questions regarding the relation between Refs.\ \onlinecite{us_dirty}
and \onlinecite{us_IFS}, and the exact critical behavior, including 
logarithmic corrections to scaling, at the quantum ferromagnetic transition 
of disordered itinerant electrons. By using the local field-theoretic 
description of I that explicitly keeps all soft modes, we show that 
Ref.\ \onlinecite{us_dirty} missed marginal operators that lead to 
logarithmic corrections to the Gaussian critical behavior discussed there. 
Moreover, taking these marginal operators into account leads to integral 
equations for the relevant vertex functions that are identical to the ones 
derived in Ref.\ \onlinecite{us_IFS}. The current formulation can further
be mapped onto the one of Ref.\ \onlinecite{us_IFS}, which shows that the
transition found in the latter paper was really the ferromagnetic one,
and that the results originally derived in that reference are exact.

This paper is organized as follows. In Sec.\ \ref{sec:II} we first recall
the results of I. We then use diagrammatic techniques to derive exact integral 
equations for the two-point vertex functions that appear in the theory. We 
conclude this section by quoting a previous solution to these equations that is 
valid at the critical point. In Sec.\ \ref{sec:III} we show how some
physical observables in the paramagnetic phase are related to these vertex
functions. We then develop a scaling theory to determine the critical
behavior of other observables of interest, as well as the critical behavior
in the ferromagnetic phase. In Sec.\ \ref{sec:IV} 
we discuss general theoretical aspects of this paper as well as
experimental consequences of our results. Various technical issues are
relegated to several appendices.

\section{Effective field theory, and its solution}
\label{sec:II}

\subsection{Effective action}
\label{subsec:II.A}

In I it was shown that the effective long-wavelength and low-frequency
field theory that contains the critical fixed point and describes the 
exact quantum critical behavior of disordered itinerant ferromagnets is 
given by the action
\bea
{\cal A}_{\rm eff}&=&-\sum_{{\bf k},n,\alpha}\sum_{i=1}^3
  {^iM}_n^{\alpha}({\bf k})\,u_2({\bf k})\,
   {^iM}_{-n}^{\alpha}(-{\bf k})
\nonumber\\
&&- \frac{4}{G}\sum_{\bf k}\sum_{1,2,3,4}\sum_{r,i}{^i_rq}_{12}({\bf k})\,
        \Gamma_{12,34}^{(2)}({\bf k})\,{^i_rq}_{34}(-{\bf k})
\nonumber\\
&&-\frac{1}{4G}\sum_{1,2,3,4}\ \sum_{r,s,t,u}\ \sum_{i_1,i_2,i_3,i_4}\frac{1}{V}
   \sum_{{\bf k}_1,{\bf k}_2,{\bf k}_3,{\bf k}_4}
\nonumber\\
&&\hskip 10pt\times {^{i_1i_2i_3i_4}_{\ \ \ rstu}\Gamma}
    ^{(4)}_{1234}({\bf k}_1,{\bf k}_2,{\bf k}_3,{\bf k}_4)\
    {^{i_1}_rq}_{12}({\bf k}_1)\,{^{i_2}_sq}_{32}({\bf k}_2)
\nonumber\\
&&\hskip 100 pt \times {^{i_3}_tq}_{34}({\bf k}_3)\,{^{i_4}_uq}_{14}({\bf k}_4)
\nonumber\\
&&+ c_1\,\sqrt{T}\sum_{\bf k}\sum_{12}\sum_{i,r}
           {^i_rb}_{12}({\bf k})\,{^i_rq}_{12}(-{\bf k})
\nonumber\\
&&+ c_2\,\sqrt{T}\,\frac{1}{\sqrt{V}}\sum_{{\bf k},{\bf p}}\sum_{n_1,n_2,m}
  \sum_{r,s,t}\sum_{i=1}^3\sum_{j,k}\sum_{\alpha,\beta}
     {^i_rb}_{n_1n_2}^{\alpha\alpha}({\bf k})\,
\nonumber\\
&&\hskip -25 pt \times\left[{^j_sq}_{n_2m}^{\alpha\beta}({\bf p})\,
  {^k_tq}_{n_1m}^{\alpha\beta}(-{\bf p}-{\bf k})\,
   \tr\left(\tau_r\tau_s\tau_t^{\dagger}\right)\,\tr\left(s_is_js_k^{\dagger}
                                                           \right)\right.
\nonumber\\
&&\hskip -20pt \left. - {^j_sq}_{mn_2}^{\beta\alpha}({\bf p})\,
  {^k_tq}_{mn_1}^{\beta\alpha}(-{\bf p}-{\bf k})\,
      \tr\left(\tau_r\tau_s^{\dagger}\tau_t\right)\,\tr\left(s_is_j^{\dagger}
                                             s_k\right)\right]\,.
\nonumber\\
\label{eq:2.1}
\eea
Here ${\bf M}_{n}^{\alpha }({\bf k})$, with components
${^iM}_{n}^{\alpha }({\bf k})$, is the fluctuating magnetization at
wavenumber ${\bf k}$ and bosonic Matsubara frequency $\Omega_{n}=2\pi Tn$, 
with $\alpha$ a replica label, and the field $b$ is defined in terms of the 
magnetization,
\bea
{^i_rb}_{12}({\bf k})&=&\delta_{\alpha_1\alpha_2} (-)^{r/2}\sum_n
     \delta_{n,n_1-n_2}\left[{^iM}_n^{\alpha_1}({\bf k})\right.
\nonumber\\
 &&\hskip 50pt \left.+ (-)^{r+1}\,{^iM}_{-n}^{\alpha_1}({\bf k})\right]\quad.
\label{eq:2.2}
\eea
The labels $1$, $2$, etc., comprise both frequency and replica indices,
$1\equiv (n_1,\alpha_1)$, etc. The two-point $M$-vertex is given by
\bml
\label{eqs:2.3}
\be
u_2({\bf k}) = t_0 + a_{d-2}\,\vert{\bf k}\vert^{d-2} + a_2\,{\bf k}^2\quad.
\label{eq:2.3a}
\ee
Here the nonanalytic term, proportional to $\vert{\bf k}\vert^{d-2}$, 
reflects the nonanalytic wavenumber dependence of the electron spin
susceptibility in a disordered itinerant electron system, as has been
explained in I. For weak disorder, characterized by a mean-free path
$\ell >> 1/k_{\rm F}$, with $k_{\rm F}$ the Fermi wavenumber, the
prefactor $a_{d-2}$ is of order $1/k_{\rm F}\ell$, while $a_2 = O(1)$.
For physical values of the spatial dimension $d$, and for asymptotically
small wavenumbers, the nonanalytic term
dominates the analytic ${\bf k}^2$-term. However, for completeness
and later reference we include the latter, which had been dropped from
the final effective action in I. $t_0$ is the bare distance from the
ferromagnetic critical point.

The fermionic degrees of freedom are represented by the field $q$;
electron number, spin, and energy density fluctuations can all be expressed
in terms of the $_{r}^{i}q_{nm}^{\alpha\beta}$. These are the additional
slow modes mentioned in Sec.\ \ref{sec:I} above. The frequency labels 
$n\geq 0$, $m<0$ of the $q$ denote fermionic Matsubara frequency indices, 
$i$ is a spin label ($i=0$ and $i=1,2,3$ correspond to spin-singlet and 
spin-triplet fluctuations, respectively), and the label $r$ ($r=0,3$) 
serves to write the complex-valued $q$ fields as 
two-component real-number valued 
fields.\cite{r_footnote} The fermionic part of the action is characterized
by the two-point vertex
\be
\Gamma _{12,34}^{(2)}({\bf k}) = \delta_{13}\delta_{24}\,({\bf k}^{2}
   + GH\Omega_{n_{1}-n_{2}})\quad,
\label{eq:2.3b}
\ee
and the four-point vertex
\bea
{^{i_1i_2i_3i_4}_{\ \ \ rstu}\Gamma}
    ^{(4)}_{1234}({\bf k}_1,{\bf k}_2,{\bf k}_3,{\bf k}_4)
  &=&-\delta_{{\bf k}_1+{\bf k}_2+{\bf k}_3+{\bf k}_4,0}\,
\nonumber\\
&&\hskip -100pt\times
 \tr\left(\tau_r\tau_s^{\dagger}\tau_t\tau_u^{\dagger}\right)\,
   \tr\left(s_{i_1}s_{i_2}^{\dagger}s_{i_3}s_{i_4}^{\dagger}
              \right)\,
     \left({\bf k}_1\cdot{\bf k}_3 + {\bf k}_1\cdot{\bf k}_4\right.
\nonumber\\
&&\hskip -75pt \left. + {\bf k}_1\cdot{\bf k}_2 +
   {\bf k}_2\cdot{\bf k}_4 - GH\Omega_{n_1-n_2} \right)\quad.
\label{eq:2.3c}
\eea
The parameter $G=8/\pi\sigma_{0}$ is a measure of the disorder, with 
$\sigma_{0}$ the bare conductivity. $H=\pi N_{F}/4$, with
$N_{F}$ the bare single-particle or tunneling density of states per spin at the 
Fermi surface, is a bare quasi-particle density of states that also
determines the specific heat coefficient. Finally, $c_1, c_2$ are coupling
constants whose bare values are related and given by
\be
c_1 = 16\, c_2 = 4\sqrt{\pi\,K_t}\quad,
\label{eq:2.3d}
\ee
\eml%
with $K_t$ the spin-triplet interaction amplitude of the electrons.
The replicated partition function is given in terms of the action by
\be
Z=\int D[{\bf M},q]\ e^{A_{\rm eff}[{\bf M},q]}\quad.
\label{eq:2.4}
\ee

\subsection{Perturbation theory to all orders}
\label{subsec:II.B}

We will now show that the effective action given in the preceding
subsection can be solved perturbatively exactly, i.e. it is possible
to resum perturbation theory to all orders.
The basic idea is to first show that the
$n$-point vertices for $n\geq 3$ are either not renormalized, or their
renormalization is simply related to that of the two-point
vertex functions. This in turn implies that exact self-consistent equations
for the two-point vertex functions can be derived. The net result will be
that the determination of the critical behavior of the field theory is
reduced to the solution of two coupled integral equations that were
first derived by different methods in Ref.\ \onlinecite{us_IFS}.

\subsubsection{Gaussian propagators}
\label{subsubsec:II.B.1}

In order to set up a loop expansion we will need the basic two-point 
propagators for the above theory. They are determined by the Gaussian action,
\bea
{\cal A}_{\rm G}[{\bf M},q]&=&-\sum_{\bf k}\sum_n\sum_{\alpha}\sum_{i=1}^3
   {^iM}_n^{\alpha}({\bf k})\,u_2({\bf k})\,{^iM}_{-n}^{\alpha}(-{\bf k})
\nonumber\\
&&\hskip -1pt - \frac{4}{G}\sum_{\bf k}\sum_{1,2,3,4}
     \sum_{i,r}{^i_rq}_{12}({\bf k})\,
        \Gamma_{12,34}^{(2)}({\bf k})\,{^i_rq}_{34}(-{\bf k})
\nonumber\\
&&+ 4\sqrt{\pi TK_t}\sum_{\bf k}\sum_{12}\sum_{i,r}{^i_rq}_{12}({\bf k})\,{^i_rb
}_{12}
      (-{\bf k})\quad.
\nonumber\\
\label{eq:2.5}
\eea
The quadratic form defined by this Gaussian action has been inverted in I. 
For the order parameter correlations we find
\bml
\label{eqs:2.6}
\be
\langle{^i\!M}_n^{\alpha }({\bf k})\,{^j\!M}_m^{\beta }({\bf p})\rangle =
   \delta_{{\bf k},-{\bf p}}\,\delta_{n,-m}\,\delta_{ij}\,\delta_{\alpha\beta}
     \,\frac{1}{2}\,{\cal M}_{n}({\bf k})\ ,
\label{eq:(2.6a)}
\ee
\bea
\langle{^i_rb}_{12}({\bf k})\,{^j_sb}_{34}({\bf p})\rangle&=&
   -\delta_{{\bf k},-{\bf p}}\,\delta_{rs}\,\delta_{ij}\,
   \delta_{\alpha_1\alpha_2}\,\delta_{\alpha_1\alpha_3}
\nonumber\\
&&\hskip -35pt\times {\cal M}_{n_1-n_2}({\bf k})\,\left[\delta_{1-2,3-4}
        - (-)^r\delta_{1-2,4-3}\right]\quad,
\nonumber\\
\label{eq:2.6b}
\eea
in terms of the paramagnon propagator,
\be
{\cal M}_n({\bf k}) = \frac{1}{t_0 + a_{d-2}\vert{\bf k}\vert^{d-2}
       + a_2\,{\bf k}^2
   + \frac{GK_t\vert\Omega_n\vert}{{\bf k}^2 + GH\vert\Omega_n\vert}}\quad.
\label{eq:2.6c}
\ee
\eml%
The dynamical piece of the paramagnon propagator ${\cal M}$, whose structure
is characteristic of disordered itinerant ferromagnets, has been produced by
the coupling between the order parameter field and the fermionic degrees of
freedom.

For the fermionic propagators one finds
\be
\langle{^i_rq}_{12}({\bf k})\,{^j_sq}_{34}({\bf p})\rangle =
   \delta_{{\bf k},-{\bf p}}\,\delta_{rs}\,\delta_{ij}\,\frac{G}{8}\,
   {^i\Gamma}_{12,34}^{(2)\,-1}({\bf k})\quad,
\label{eq:2.7}
\ee
in terms of the inverse of $\Gamma^{(2)}$, 
\bml
\label{eqs:2.8}
\be
{^0\Gamma}_{12,34}^{(2)\,-1}({\bf k}) = \delta_{13}\,\delta_{24}\,
   {\cal D}_{n_1-n_2}({\bf k})\quad,
\label{eq:2.8a}
\ee
and the propagator
\bea
{^{1,2,3}\Gamma}_{12,34}^{(2)\,-1}({\bf k})&=&\delta_{13}\,\delta_{24}\,
   {\cal D}_{n_1-n_2}({\bf k}) - \delta_{1-2,3-4}\,\delta_{\alpha_1\alpha_2}
      \delta_{\alpha_1\alpha_3}\,
\nonumber\\
&&\times 2\pi TGK_t\,\left({\cal D}_{n_1-n_2}({\bf k})
        \right)^2\,{\cal M}_{n_1-n_2}({\bf k})\ .
\nonumber\\
\label{eq:2.8b}
\eea
\eml%
Here ${\cal D}$ is the basic diffusion propagator or diffuson. In the limit 
of small frequencies and wavenumbers it reads
\be
{\cal D}_n({\bf k}) = \frac{1}{{\bf k}^{2}+GH\Omega _{n}}\quad.
\label{eq:2.9}
\ee
Physically, ${\cal D}$ describes heat diffusion.\cite{CD,us_R}

Finally, due to the coupling between $M$ and $q$ there is a mixed propagator,
\bea
\langle{^i_rq}_{12}({\bf k})\,{^j_sb}_{34}({\bf p})\rangle&=&
   -\delta_{{\bf k},-{\bf p}}\,\delta_{rs}\,\delta_{ij}\,
   \delta_{\alpha_1\alpha_2}\,\delta_{\alpha_1\alpha_3}\,\frac{G}{2}\,
\sqrt{\pi TK_t}\,
\nonumber\\
&&\hskip -70pt {\cal D}_{n_1-n_2}({\bf k})\,{\cal M}_{n_1-n_2}({\bf k})\,
   \left[\delta_{1-2,3-4} + (-)^{r+1}\delta_{1-2,4-3}\right] \, .
\nonumber\\
\label{eq:2.10}
\eea

\subsubsection{$3$-point and $4$-point vertices}
\label{subsubsec:II.B.2}

\begin{figure}[t,h]
\centerline{\psfig{figure=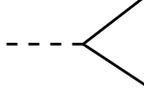,width=20mm}\vspace*{5mm}}
\caption{Diagrammatic representation of the bare $b\,q^2$-vertex. Dashed lines
 denote $M$ or $b$-fields, and solid lines denote $q$-fields.} 
\label{fig:1}
\end{figure}

\begin{figure}[t,h]
\centerline{\psfig{figure=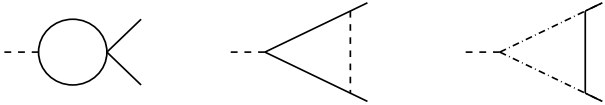,width=80mm}\vspace*{5mm}}
\caption{One-loop corrections to the vertex shown in Fig.\ \ref{fig:1}.
 Solid lines denote $\langle qq\rangle$ propagators, dashed lines 
 $\langle bb\rangle$ propagators, and dashed-dotted lines
 $\langle bq\rangle$ propagators.}
\label{fig:2}
\end{figure}

We now determine loop corrections to the tree-level theory.
We begin by considering the three-point vertex whose coupling constant
is denoted by $c_{2}$ in Eq.\ (\ref{eq:2.1}). 
Diagrammatically the bare three-point vertex is given in Fig.\ \ref{fig:1}.
Now consider the one-loop renormalizations of this vertex, which are shown in 
Fig.\ \ref{fig:2}.
For scaling purposes, we can use simple estimates for the propagators.
Specifically, the $q$-propagators all scale like an inverse wavenumber
squared,\cite{notation_footnote}
\bml
\label{eqs:2.11}
\be
\langle q_{12}({\bf k})\,q_{34}(-{\bf k})\rangle \sim 1/{\bf k}^{2}\quad.
\label{eq:2.11a}
\ee
The $M$-propagator at criticality may scale like a number, or like an
inverse wavenumber to the power $d-2$. This depends on the scaling behavior
of the frequency, which can be different in different contexts, as has been
explained in I and can be seen from Eqs.\ (\ref{eq:2.6c}) and (\ref{eq:2.9}),
respectively: $\Omega$ can scale either like $\vert{\bf k}\vert^d$, as in
the paramagnon propagator, or like ${\bf k}^2$, as in the diffuson. 
The two possibilities therefore are,
\be
\langle M_1({\bf k})\,M_2(-{\bf k})\rangle \sim 
                      \cases{{\rm const.} &if $\quad\Omega\sim {\bf k}^2$\cr
                1/\vert{\bf k}\vert^{d-2} &if $\quad\Omega\sim \vert{\bf k}
                                                           \vert^d$\cr}\quad.
\label{eq:2.11b}
\ee
Similarly, the mixed propagator, Eq.\ (\ref{eq:2.10}), scales like
\be
\langle q_{12}({\bf k})\,b_{34}(-{\bf k})\rangle \sim
             \cases{1/\vert{\bf k}\vert &if $\quad\Omega\sim {\bf k}^2$\cr
                1/\vert{\bf k}\vert^{d/2} &if $\quad\Omega\sim \vert{\bf k}
                                                           \vert^d$\cr}\quad.
\label{eq:2.11c}
\ee
\eml%
If we use an infrared wavenumber cutoff $\Lambda$, we see that the integrals
that correspond to the diagrams shown in Fig.\ \ref{fig:2} all scale like
$\Lambda^{d-2}$. That is, the one-loop renormalization of $c_2$ at zero 
external wavenumber and frequency is a finite number for all $d>2$.
More generally, an $n$-loop skeleton diagram has $n$ independent wavenumber 
and frequency integrals. Diagrams that contain only solid and dashed lines
contain $2n$ $\langle qq\rangle$ propagators, and up to $n$ 
$\langle MM\rangle$ propagators. Similar considerations hold for diagrams
that contain mixed propagators. The net result is that any $n$-loop 
skeleton diagram scales like $\Lambda^{n(d-2)}$. All of these contributions 
thus amount to finite corrections to the bare value of $c_2$. By induction
it follows that insertions do not produce singular contributions either. 
We conclude that there are no singular renormalizations of the three-point 
vertex function in the field theory defined by Eq.\ (\ref{eq:2.1}).

In addition to the renormalization of $c_2$, a new three-point vertex with a
replica structure that is different from the one of $c_2$ is 
generated by the renormalization group at one-loop order.
As shown in Appendix\ \ref{app:A}, the frequency structure of this vertex 
is such that
it carries one more frequency sum and associated temperature
factor than the vertex with coupling constant $c_2$. Its coupling constant
is therefore more irrelevant than $c_2$ and can be neglected.

\begin{figure}[t,h]
\centerline{\psfig{figure=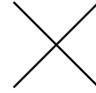,width=12mm}\vspace*{5mm}}
\caption{The bare four-point vertex}
\label{fig:3}
\end{figure}

\begin{figure}[t,h]
\centerline{\psfig{figure=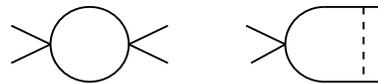,width=50mm}\vspace*{5mm}}
\caption{One-loop corrections to the vertex shown in Fig.\ \ref{fig:3}.}
\label{fig:4}
\end{figure}

Next we consider the four-point vertex, $\Gamma^{(4)}$ in Eq.\ (\ref{eq:2.1}).
The bare four-point vertex is given analytically in Eq.\ (\ref{eq:2.3c})
and shown diagrammatically in Fig.\ \ref{fig:3},
and the one-loop
renormalizations are shown in Fig.\ \ref{fig:4}.
Notice that this vertex is proportional either to a wavenumber squared, 
or to a frequency. Using the estimates given by Eqs.\ (\ref{eqs:2.11}), 
we find that the renormalization of the part of $\Gamma^{(4)}$ that is 
proportional to a wavenumber squared is always a finite number, again 
scaling as $\Lambda^{(d-2)}$. For the part that is proportional to
frequency, on the other hand, power counting shows that this term can 
have logarithmically singular renormalizations. An explicit calculation
would be very cumbersome. However, it is not necessary since the one-loop 
renormalization 
of the coupling constant $H$ in Eq.\ (\ref{eq:2.3c}) obtained this way
is identical to that obtained by renormalizing the two-point vertex,
Eq.\ (\ref{eq:2.3b}). This is because both terms arise
from the same term in the underlying nonlinear sigma model for the 
fermionic degrees of freedom, which is believed to be 
renormalizable.\cite{renormalizability_footnote} By the same argument,
the renormalization of $\Gamma^{(4)}$ to {\em all orders} is given by
that of $\Gamma^{(2)}$, and therefore need not be considered separately.
The explicit calculation of $\Gamma^{(2)}$ confirms the existence of
the logarithms that were alluded to above, as we will demonstrate in
the next subsection.

In addition to the diagrams that renormalize $\Gamma^{(4)}$, there are
one-loop terms that represent four-point vertices with more restrictive
replica structures. These correspond either to the two-body interaction
terms that were shown in I to not change the critical behavior, or to
many-body interactions that are shown in Appendix \ref{app:A} to be
more irrelevant than $\Gamma^{(4)}$ and thus can be neglected.

\subsubsection{$2$-point vertices}
\label{subsubsec:II.B.3}

\begin{figure}[t,h]
\centerline{\psfig{figure=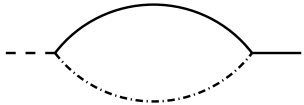,width=40mm}\vspace*{5mm}}
\caption{One-loop renormalizations of the mixed two-point vertex.}
\label{fig:5}
\end{figure}

We now turn to the two-point vertices in the effective action.
The one-loop renormalization of the $bq$-vertex is shown in
Fig.\ \ref{fig:5}.
Using Eqs.\ (\ref{eqs:2.11}), it is easy to see that this diagram is
finite in $d>2$. Since the three-point vertex is not singularly
renormalized, see the previous subsection, it follows that the
$bq$-vertex has only finite renormalizations to all orders in
perturbation theory. This means that the coupling constant $K_t$ is
not singularly renormalized.

\begin{figure}[t,h]
\centerline{\psfig{figure=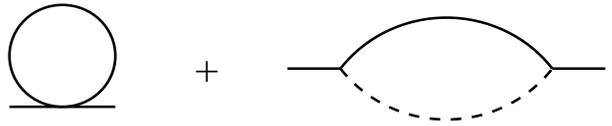,width=80mm}\vspace*{5mm}}
\caption{One-loop renormalizations of the fermionic two-point vertex 
 $\Gamma^{(2)}$.}
\label{fig:6}
\end{figure}

The one-loop renormalizations of $\Gamma^{(2)}$ are shown in 
Fig.\ \ref{fig:6}.
As was shown in I, the renormalization of $G$ obtained from these
diagrams is finite in $d>2$. For the renormalized $H$, which we denote
by $H(i\Omega_n)$, one obtains 
to one-loop order,
\be 
H(i\Omega_n) = H + \frac{3}{8}\,GK_t\,\frac{2\pi T}{\Omega_n}
              \sum_{l=0}^{n}\frac{1}{V}\sum_{\bf p}{\cal D}_l({\bf p})\,
              {\cal M}_l({\bf p})\quad,
\label{eq:2.12}
\ee 
which diverges logarithmically as $\Omega_n\rightarrow 0$ for all $2<d<4$. 
As was explained
in I, this divergent renormalization, which arises from the nominally
irrelevant vertices $\Gamma^{(4)}$ and $c_2$, is a consequence of the
presence of two time scales in the problem. In addition,
there are terms that are finite in $d>2$. It is important to note that
the structure ${\cal D}{\cal M}$ in the integrand of Eq.\ (\ref{eq:2.12})
stems from the second term, proportional to ${\cal D}^2{\cal M}$, 
in the triplet $qq$-propagator, 
Eq.\ (\ref{eq:2.8b}), times a term ${\cal D}^{-1}$ that is due to the
wavenumber and frequency dependence of the quartic vertex, 
Eq.\ (\ref{eq:2.3c}).

\begin{figure}[t,h]
\centerline{\psfig{figure=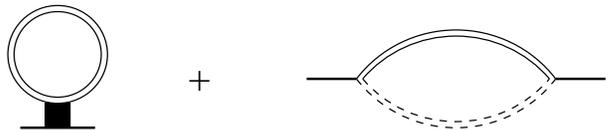,width=80mm}\vspace*{5mm}}
\caption{Renormalization of the fermionic two-point vertex $\Gamma^{(2)}$
 to all orders.}
\label{fig:7}
\end{figure}

\begin{figure}[t,h]
\centerline{\psfig{figure=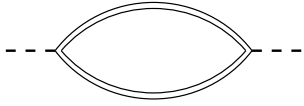,width=40mm}\vspace*{5mm}}
\caption{Renormalization of the two-point magnetization vertex $u_2$
 to all orders.}
\label{fig:8}
\end{figure}

The exact vertex $\Gamma^{(2)}$ can be generated from the one-loop diagrams
by dressing all propagators and all vertices in Fig.\ \ref{fig:6}. The
relevant vertices are $\Gamma^{(4)}$ and $c_2$. As was shown in
Sec.\ \ref{subsubsec:II.B.2}, the latter has no singular renormalizations
in $d>2$, so it need not be dressed. Denoting the exact four-$q$ vertex
by a square, and the dressed propagators by double lines, we therefore
have the diagrammatic representation of the renormalization of $\Gamma^{(2)}$
to {\em all orders} shown in Fig.\ \ref{fig:7}.
Analytically, this result corresponds to simply dressing the propagators
in Eq.\ (\ref{eq:2.12}). Notice that this procedure includes the vertex
renormalization due to the structure pointed out above. Also notice
that it is crucial for our argument that $G$ and $K_t$ carry finite
renormalizations only. We thus have the exact result, as far as the
asymptotic critical behavior is concerned,
\bml
\label{eqs:2.13}
\bea
H(i\Omega_n)&=&H + \frac{3}{8}\,GK_t\,\frac{2\pi T}{\Omega_n}\sum_{l=0}^{n}
              \frac{1}{V}\sum_{\bf p}
\nonumber\\
&&\hskip 30pt\times\frac{1}{GK_t\Omega_l + {\bf p}^2u_2({\bf p},i\Omega_l)}
   \quad,
\label{eq:2.13a}
\eea
with $u_2({\bf p},i\Omega_l)$ the fully renormalized $bb$-vertex.
By the same arguments, we obtain the latter as shown in Fig.\ \ref{fig:8}.
By dressing the propagators in the analytic one-loop expression given
in Eq. (3.5b) of I, we have, for $\Omega_n\geq 0$,
\bea
u_2({\bf k},i\Omega_n)&=&t_0 - \frac{G^2}{2}\,\frac{K_t}{V}\sum_{\bf p}
        2\pi T\sum_{l=0}^{\infty}\frac{1}{{\bf p}^2 + GH(i\Omega_l)\,\Omega_l}\,
\nonumber\\
&&\hskip -31pt \times\frac{1}{({\bf p}+{\bf k})^2 + GH(i\Omega_l+i\Omega_n)
     (\Omega_l+\Omega_n)}\quad.
\label{eq:2.13b}
\eea
\eml%
In writing Eq.\ (\ref{eq:2.13b}), we have for simplicity put the bare 
coupling constants $a_{d-2}$ and $a_2$ equal to zero, since they are 
generated at one-loop order.

\subsection{Integral equations for diffusion coefficients}
\label{subsec:II.C}

Eqs.\ (\ref{eqs:2.13}) in Sec.\ \ref{subsec:II.B} constitute two closed
integral equations for the two-point vertices. As we have seen, this has
been possible to achieve since (1) the four-point vertex $\Gamma^{(4)}$
renormalizes like the two-point vertex $\Gamma^{(2)}$, and (2) all other
vertices are subject to finite renormalizations only. As a result, the
solution of Eqs.\ (\ref{eqs:2.13}) provides us with the perturbatively
{\em exact} critical behavior.

To make contact with previous work, it is useful to rewrite 
Eqs.\ (\ref{eqs:2.13}) in terms of
the (thermal) diffusion coefficient $D(i\Omega_n) = 1/GH(i\Omega_n)$
and the spin diffusion coefficient, 
$D_{s}({\bf k},i\Omega_n) = u_{2}({\bf k},i\Omega_n)/GK_t$.
If we analytically continue to real 
frequencies, $i\Omega_n\rightarrow \Omega + i0$, 
the self-consistent one-loop equations read
\bml
\label{eqs:2.14}
\bea
D_{s}({\bf k},\Omega)&=&D_{s}^{0} + \frac{iG}{2V}\sum_{\bf p}
   \int_{0}^{\infty }d\omega \frac{1}{{\bf p}^{2} - i\omega/D(\omega)}
\nonumber\\
&&\times\frac{1}{({\bf p+k)}^{2}-i(\omega +\Omega)/D(\omega +\Omega )}\quad,
\label{eq:2.14a}
\eea
\be
\frac{1}{D(\Omega)} = \frac{1}{D^{0}} + \frac{3G}{8V}\sum_{{\bf p}}
   \frac{1}{\Omega} \int_{0}^{\Omega}d\omega \frac{1}{-i\omega 
   + {\bf p}^{2}D_{s}({\bf p},\omega )}\quad.
\label{eq:2.14b}
\ee
\eml%
Here $D^0 = 1/GH$ and $D_s^0 = t_0/GK_t$ are the bare diffusion coefficients.
These integral equations were first derived in Ref. \onlinecite{us_IFS}
by means of a resummation of perturbation theory within a nonlinear sigma
model for interacting electrons in the limit of a large spin-triplet 
interaction amplitude. As has been discussed in I, this sigma model is
recovered from the current model by integrating out the magnetization.
This is the mapping between the two models that was referred to in the 
Introduction.

\subsection{Solution of the integral equations}
\label{subsec:II.D}  

In Ref.\ \onlinecite{us_IFS}, the coupled integral equations, 
Eqs.\ (\ref{eqs:2.14}), were solved by
three distinct methods: A direct analytic solution, a renormalization group
solution, and a numerical solution. The results of all three 
approaches were consistent with one another. Here we will quote the most
relevant results, restricting ourselves, as in I, to $2<d<4$.

Simple scaling arguments show that at criticality, 
$t\equiv u_2({\bf k}=0,\Omega_n=0) = 0$, 
$D(\Omega\rightarrow 0)$ is a constant except for logarithmic terms. 
This suggests the {\em ansatz}
\be
D(\Omega) = D^{0}/F[\ln (1/\Omega \tau )].
\label{eq:2.15}
\ee
Here $1/\tau =\pi nG/8m$, with $n$ the electron density and $m$ the electron
effective mass, is the elastic-scattering rate. The same arguments yield 
$D_{s}({\bf k},\Omega=0)\sim \vert{\bf k}\vert^{d-2}$ at the critical point 
except for logarithmic terms. So we write,
\be
D_{s}({\bf k},\Omega=0) = D_{s}^{0}\,(\vert{\bf k}\vert/k_{F})^{d-2}\,
   F_{s}[\ln (k_{F}/\vert{\bf k}\vert)]\quad.
\label{eq:2.16}
\ee
Solving the resulting equations for 
$F$ and $F_{s}$ gives\cite{us_IFS}
\bml
\label{eqs:2.17}
\be
D(\Omega \rightarrow 0) = D^{0}\,\left[g(\ln 1/\Omega\tau))\right]^{-1}\quad,
\label{eq:2.17a}
\ee
\bea
D_{s}({\bf k}\rightarrow 0,\Omega=0)&=&D_{s}^{0}\,
   (\vert{\bf k}\vert/k_{F})^{d-2}\,d'\,(GK_t/H)\,k_{F}^{d-2}
\nonumber\\
&&\times\left[g(2\ln (k_{\rm F}/\vert{\bf k}\vert))\right]^{-1}\quad.
\label{eq:2.17b}
\eea
\eml%
Here
\bml
\label{eqs:2.18}
\be
g(x) = \sum_{n=0}^{\infty} [(c(d)\,x)^n/n!]\,e^{(n^2-n)\ln(2/d)/2}\quad.
\label{eq:2.18a}
\ee
In an asymptotic expansion for large $x$, the leading term is
\be
g(x) \approx \left[2\ln (d/2)/\pi\right]^{-1/2}\,e^{[\ln (c(d)\,x)]^2/
       2\ln (d/2)}\quad.
\label{eq:2.18b}
\ee
\eml%
The dimensionality dependent coefficient $c(d)$ is given by
\bml
\label{eqs:2.19}
\be
c(d)=c'(d)/d'(d)\quad,
\label{eq:2.19a}
\ee
where
\be
d'(d) = c''\Gamma (2-d/2)\,F(2-d/2,1/2;3/2;1)\quad,
\label{eq:2.19b}
\ee
\eml%
with $F$ a hypergeometric function, $\Gamma $ the gamma function, and $c'$ 
and $c''$ smoothly varying functions of $d$.

The leading dependence of either $D$ or $D_s$ on $t$ away from criticality at
zero frequency and wavenumber, or that of $D_s$ at $t=0$ as a function of 
frequency at ${\bf k}=0$, will follow from the scaling theory to be
developed in Sec.\ \ref{sec:III} below.

\section{Critical behavior of observables}
\label{sec:III}

In this section we determine the exact critical behavior of various
observables. We do so by developing a general scaling description for
the free energy and various transport coefficients and relaxation
rates, and using the exact solution of Sec.\ \ref{sec:II} to determine
the values of the independent critical exponents. Again, we
restrict ourselves to the dimensionality range $2<d<4$.
The results for $d>4$ in Ref.\ \onlinecite{us_dirty} were exact, and the 
behavior in $d=4$ can be obtained by combining the solution of the
integral equations from Ref.\ \onlinecite{us_IFS} for that case with
the arguments given below. We have also checked some of our results by
means of explicit perturbation theory, see Appendix \ref{app:B}. 
Parts of the results presented here have been
previously published in Ref.\ \onlinecite{us_letter}.

\subsection{Identification of observables}
\label{subsec:III.A}

We first discuss how to relate the behavior of some physical 
observables of interest near the quantum critical point to the solution of
the effective field theory given in Sec.\ \ref{sec:II}.
We consider the electrical conductivity $\sigma$,
the specific heat coefficient $\gamma_C$, the tunneling density of states $N$,
the spin susceptibility $\chi_s$, the heat and spin diffusion coefficients
$D$ and $D_s$, respectively, the phase relaxation rate $\tau_{\rm ph}^{-1}$, 
and various quasiparticle properties, in particular the quasiparticle
decay rate $\tau_{\rm QP}^{-1}$. The magnetization $m$ will be obtained
from the free energy in Sec.\ \ref{subsec:III.B} below.

The conductivity, $\sigma = 8/\pi G = 1/\rho$, with $\rho$ the resistivity,
is proportional to the inverse of the renormalized disorder parameter, $G$, in 
Eq.\ (\ref{eq:2.1}), and the specific heat coefficient, $\gamma_C = C/T$, 
with $C$ the specific heat, is proportional to the
renormalized value of $H$ in Eqs.\ (\ref{eq:2.3b},\ref{eq:2.3c}).\cite{us_R}
The single particle density of states (DOS) as a function of the distance
in energy or frequency space from the Fermi surface, N($\epsilon)$, is given by 
Eq.\ (2.29d) in I. In terms of expectation values of the field $q$, it
takes the form of an expansion
\bea
N(\epsilon_{F} + \epsilon)&=&N_{F}\left[1 - \frac{1}{2}\sum_{\beta}\sum_m
   \sum_{i,r}\langle {^i_rq}_{nm}^{\alpha\beta}({\bf x})\,\right.
\nonumber\\
&&\hskip 10pt \times
         {^i_rq}_{nm}^{\alpha\beta}({\bf x})
          \rangle_{i\omega_{n}\rightarrow \epsilon +i0} 
     + O(\langle q^4\rangle)\biggr]\quad.
\label{eq:3.1}
\eea
Experimentally, $\epsilon$ is equal to the electron charge times the
bias voltage.
The dynamical spin susceptibility is given by the ${\bf M}\cdot{\bf M}$
correlation function, and we have seen that the coupling between the
${\bf M}$ and $q$ fields affects the dynamical part of that correlation
function only. This implies that the static spin susceptibility,
$\chi_{s}({\bf k})$, is given by the renormalized value of 
$u_{2}({\bf k},i\Omega_n=0)$, cf. Eq.  (\ref{eq:2.13b}). Alternatively, 
it is proportional to the inverse of the static spin diffusion coefficient,
\be
\chi _{s}({\bf k})=\frac{1}{2GK_{t}D_{s}({\bf k},i\Omega_n=0)}\quad.
\label{eq:3.2}
\ee
The heat and spin diffusion coefficients $D$ and $D_{s}$ are 
explicitly given by the solution of the integral equations discussed in
Sec.\ \ref{sec:II}. In Appendix\ \ref{app:C} we 
show how to generalize the integral equations 
to the ordered phase, so that the equation of state or the magnetization, $m$, 
can be obtained. In Sec.\ \ref{subsec:III.B} we determine the magnetization
by means of a scaling theory.

The relaxation rates are defined in terms of the diffuson, whose bare
propagator is given by Eq.\ (\ref{eq:2.9}). Its renormalized counterpart
has the form
\bml
\label{eqs:3.3}
\be
{\cal D}({\bf k},i\Omega_n) = \frac{Z^2}{{\bf k}^2 + GH(i\Omega_n)\Omega_n}
  = \frac{Z^2}{{\bf k}^2 + \Omega_n/D(i\Omega_n)}
   \quad.
\label{eq:3.3a}
\ee
Here $Z$ is the wavefunction renormalization, which determines the 
single-particle DOS, Eq. (\ref{eq:3.1}), via $N = N_{\rm F}Z$. In
addition, the quasiparticle DOS $N_{\rm QP}$ is related to $N$ via
\be
N_{\rm QP} = N/a_{\rm QP}\quad,
\label{eq:3.3b}
\ee
with $a_{\rm QP}$ the quasiparticle weight. Upon an analytic continuation to
real frequencies, $H(i\Omega_n)$ 
acquires a real part $H'$ and an imaginary part $H''$,
$H(i\Omega_n\rightarrow\epsilon + i0) = H'(\epsilon) + iH''(\epsilon)$. Dividing
by $GH'$, the renormalized diffuson can be written
\be
{\cal D}({\bf k},i\Omega_n\rightarrow\epsilon+i0) 
  = \frac{1}{G}\,\frac{(N_{\rm QP}/N_{\rm F}^2)\,a_{\rm QP}^2}
   {D_{\rm QP}{\bf k}^2 - i\epsilon + \tau_{\rm QP}^{-1}}\quad,
\label{eq:3.3c}
\ee
\eml%
with $a_{\rm QP} = N_{\rm F}Z/H'(\epsilon)$, $N_{\rm QP} = H'(\epsilon)$, 
$D_{\rm QP} = 1/GH'(\epsilon)$ the quasiparticle diffusion coefficient,
and
\bml
\label{eqs:3.4}
\be
\tau_{\rm QP}^{-1} = \epsilon H''(\epsilon)/H'(\epsilon) \quad,
\label{eq:3.4a}
\ee
the quasiparticle decay rate. In contrast, 
the phase breaking rate $\tau_{\rm ph}^{-1}$ is the ``mass'' that is
acquired by the diffuson at real frequencies and
finite temperature. Equation (\ref{eq:3.3a}) shows that it can be
identified, apart from a multiplicative constant, with
\be
\tau_{\rm ph}^{-1} = {\rm Re}(\Omega H(i\Omega))/N_{\rm F}
  \Bigl\vert_{i\Omega\rightarrow\epsilon+i0}\quad,
\label{eq:3.4b}
\ee
or, in terms of the imaginary part of $H$,
\be
\tau_{\rm ph}^{-1} = \epsilon H''(\epsilon)/N_{\rm F}\quad.
\label{eq:3.4c}
\ee
\eml%

For later reference we also give the renormalized paramagnon propagator.
Leaving out terms that are irrelevant for our purposes, it reads
\bea
{\cal M}({\bf k},i\Omega_n)&=&\frac{1}{u_2({\bf k},i\Omega_n) + 
   GK_t\vert\Omega_n\vert/{\bf k}^2}
\nonumber\\
&=&\frac{1/GK_t}{D_s({\bf k},i\Omega_n) + \vert\Omega_n\vert/{\bf k}^2}\quad.
\label{eq:3.5}
\eea

The above quantities have all been defined at $T=0$, with an eye on the
fact that the results of Sec.\ \ref{sec:II} are only valid at $T=0$.
In general, we are also interested in the analogous results at $T>0$.
This is of particular importance for the relaxation rates, since only
at $T>0$ do they provide a mass for the diffuson propagator and hence
constitute a true real-time decay rate. In order to obtain complete
results at finite temperatures, a Matsubara frequency sum
and an analytic continuation to real frequencies need to be performed.
The leading temperature dependence of the inelastic
scattering rates actually comes from a branch cut in this analytic
continuation.\cite{FA} It turns out that 
to capture this effect one needs to
retain terms that were neglected in the derivation of the
integral equations in Sec.\ \ref{sec:II}. Alternatively, the
inelastic scattering rates at $T=0$ as functions of real frequency,
or the distance in energy space from the Fermi surface, 
can be obtained from the results of Sec. \ref{sec:II}, see
Eqs.\ (\ref{eqs:3.4}) above.
Scaling theory can then be used to obtain the results at $\epsilon=0$ and
$T>0$. We will follow this second route.

\subsection{Scaling Considerations}
\label{subsec:III.B}

In this section we develop a general scaling theory for the physical
observables near the quantum critical point. The explicit solution given
in Sec. \ref{subsec:II.D} is used to identify the results for the 
critical exponents.

\subsubsection{Critical exponents}
\label{subsubsec:III.B.1}

From the explicit solution quoted in Sec.\ \ref{sec:II} above, we 
obtain various critical
exponents. As has been pointed out in I, there are multiple dynamical
exponents. This is also obvious from Eqs.\ (\ref{eqs:2.14}) in conjunction
with Eqs.\ (\ref{eqs:2.17}). Furthermore, the logarithmic corrections to
scaling that characterize the solution of the field theory mean that the
asymptotic critical behavior is {\rm not} given by simple power laws.
A convenient way to account for that is to write the critical behavior
as power laws with scale dependent critical exponents. For instance,
the critical time scale, which determines the dynamics of the paramagnon
propagator, is given by a dynamical exponent
\bml
\label{eqs:3.6}
\be z_c = d + \ln g(\ln b)/\ln b\quad,
\label{eq:3.6a}
\ee
with $b$ an arbitrary renormalization group length scale factor.
To see this, consider Eq.\ (\ref{eq:3.5}) in conjunction with
Eq.\ (\ref{eq:2.17b}), which show that the frequency in the paramagnon
scales like $\Omega\sim {\bf k}^2 D_s({\bf k},\Omega=0)\sim
\vert{\bf k}\vert^d/g(\ln(k_{\rm F}/\vert{\bf k}\vert))$.
In addition, there are diffusive time scales with power $2$ and various
logarithmic corrections. For instance, the frequency $\Omega$ in the
dressed diffuson, Eq.\ (\ref{eq:3.3a}), defines a time scale with
a critical exponent
\be
{\tilde z}_{\rm d} = 2 + \ln g(\ln b)/\ln b\quad.
\label{eq:3.6b}
\ee
There are other diffusive time scales, however. For instance,
the quantity $H(i\Omega_n)\,\Omega_n/N_{\rm F}$ is dimensionally
a frequency that defines a time scale with critical exponent
\be
z_{\rm d} = 2\quad.
\label{eq:3.6c}
\ee
This is the scale dimension of the phase relaxation rate,
Eq.\ (\ref{eq:3.4b}).
For the other two independent critical exponents, we pick $\eta$, which
describes the wavenumber dependence of the order parameter susceptibility,
and the correlation length exponent $\nu$. From Eq.\ (\ref{eq:2.17b}),
we have for the former
\be
\eta = 4 - d - \ln g(\ln b)/\ln b\quad.
\label{eq:3.6d}
\ee
The exponent $\nu$ was determined in Ref.\ \onlinecite{us_IFS} with
the result
\be
1/\nu = d - 2 + \ln g(\ln b)/\ln b\quad.
\label{eq:3.6e}
\ee
\eml%
The scale dependent exponents shown determine the asymptotic critical
behavior, including the leading logarithmic corrections to scaling.
They do not include logarithmic terms that are less leading than the
log-log-normal dependence due to the function $g(\ln b)$. In particular,
simple powers of logarithms would correspond to terms of order 
$\ln\ln b/\ln b$ in Eqs.\ (\ref{eqs:3.6}).

\subsubsection{Thermodynamic quantities, and the density of states}
\label{subsubsec:III.B.2}

We start by considering the
thermodynamic properties near the phase transition. They can all be
obtained by a scaling ansatz for the free energy as a function of $t$, $T$,
and $h$, with $h$ the magnetic field. Two key ideas will be used. First, the
existence of two essentially different time scales, see 
Sec.\ \ref{subsubsec:III.B.1}, implies that the free
energy density, $f$, should consist of two scaling parts. The second idea
is that $h$ also represents an energy scale (namely, the Zeeman energy)
and it therefore scales like the frequency or the temperature.

Taking all of this into account, the natural scaling ansatz for $f$ is
\bea
f(t,T,h)&=&b^{-(d+z_c)}\,f_{1}(tb^{1/\nu},Tb^{z_c},hb^{z_c})
\nonumber\\
&&\hskip -20pt +b^{-(d+{\tilde z}_{\rm d})}\,
         f_{2}(tb^{1/\nu},Tb^{{\tilde z}_d},hb^{z_c})\quad,
\label{eq:3.7}
\eea
with $f_1$ and $f_2$ scaling functions. 
Note that {\it a priori} there is no reason for the temperature argument 
of $f_{2}$ to be given by the diffusive time scale with the dynamical
exponent ${\tilde{z}_d}$ rather than by the critical time scale with
the dynamical exponent $z_c$. It requires explicit calculations to see
that for some quantities, e.g., the magnetization, the diffusive temperature 
scale is the relevant one. In Appendix\ \ref{app:C} we show that the equation 
of state contains only the diffusive temperature scale, and our scaling ansatz 
for $f_{2}$ reflects this feature (see also Ref. \onlinecite{us_dirty}).
In addition to the $T$ and $h$
dependences in Eq. (\ref{eq:3.7}) with scales determined by $z_c$, 
there are subleading dependences involving the diffusive time scale that we
suppress. The magnetization $m$, specific heat $C$, and spin susceptibility
$\chi_s$, respectively, are given by
\bml
\label{eqs:3.8}
\bea
m&=&\partial f/\partial h\quad,
\label{eq:3.8a}\\
C&=&-T\partial^2 f/\partial T^2\quad,
\label{eq:3.8b}\\
\chi_s&=&\partial^2 f/\partial h^2\quad.
\label{eq:3.8c}
\eea
\eml%
Generalized homogeneity laws for $m$, $\gamma_C = C/T$ and $\chi _{s}$ are
obtained by using Eq. (\ref{eq:3.7}) in Eqs.\ (\ref{eqs:3.8}). Substituting
the exponent values given by Eqs.\ (\ref{eqs:3.6}), we find
\bml
\label{eqs:3.9}
\bea
m(t,T,h)&=&b^{-2}f_{m}(t b^{d-2} g(\ln b),T b^2 g(\ln b),h b^d g(\ln b))\ ,
\nonumber\\
\label{eq:3.9a}\\
\gamma_C (t,T,h)&=&g(\ln b)\,
\nonumber\\
&&\times f_{\gamma }(t b^{d-2} g(\ln b), T b^d g(\ln b),
                 h b^d g(\ln b))\quad,
\nonumber\\
\label{eq:3.9b}\\
\chi_{s}(t,T,h)&=&b^{d-2}\,g(\ln b)\,
\nonumber\\
&&\times f_{\chi}(t b^{d-2} g(\ln b), Tb^2 g(\ln b), h b^d g(\ln b))\ ,
\nonumber\\
\label{eq:3.9c}
\eea
\eml%
with $f_m$, $f_{\gamma}$, and $f_{\chi}$ scaling functions.
With suitable choices of the scale factor $b$, Eqs.\ (\ref{eqs:3.9}) imply
\bml
\label{eqs:3.10}
\bea
m(t,0,0)&=&f_{m}(1,0,0)\,\left[t\,g\left(\frac{1}{d-2}\ln\frac{1}{t}\right)
           \right]^{2/(d-2)}\quad,
\nonumber\\
\label{eq:3.10a}\\
m(0,0,h)&=&f_{m}(0,0,1)\,\left[h\,g\left(\frac{1}{d}\ln\frac{1}{h}\right)
           \right]^{2/d}\quad,
\nonumber\\
\label{eq:3.10b}\\
\gamma_C (0,T,0)&=&f_{\gamma }(0,1,0)\,g\left(\frac{1}{d}\ln\frac{1}{T}\right)
    \quad,
\label{eq:3.10c}\\
\chi_{s}(t,0,0)&=&f_{\chi}(1,0,0)\,t^{-1}\quad,
\label{eq:3.10d}
\eea
\eml%
for the leading asymptotic critical behavior in the sense explained at
the end of Sec.\ \ref{subsubsec:III.B.1}.
For the critical exponents $\beta$, $\gamma$, $\delta$, and $\alpha$,
defined by $m\propto t^{\beta}$, $\chi_s\propto t^{-\gamma}$,
$m\propto h^{1/\delta}$, and $\gamma_C\propto T^{-(1+\alpha)}$ (the latter
is a generalization to zero-temperature transitions of the usual definition
of the exponent $\alpha$), we obtain from Eqs.\ (\ref{eqs:3.10})
\bml
\label{eqs:3.11}
\bea
\beta &=& 2\nu\quad,
\label{eq:3.11a}\\
\gamma &=& 1\quad,
\label{eq:3.11b}\\
\delta &=& z_c/2\quad,
\label{eq:3.11c}\\
\alpha &=& -d/z_c\quad,
\label{eq:3.11d}
\eea
\eml%
with $\nu$ and $z_c$ from Eqs.\ (\ref{eqs:3.6}).

Next we consider the single-particle density of states. We define
$\Delta N = N - N_{F}$, where $N_{\rm F}$ is the disordered Fermi liquid
value of $N$. $\Delta N$ can be related to a correlation function
that has scale dimension $-(d-2)$.\cite{us_fermions} This implies
the scaling form
\bml
\label{eqs:3.12}
\bea
\Delta N(t,\epsilon,T)&=&b^{-(d-2)}
\nonumber\\
&&\hskip -45pt \times f_{N}(t b^{d-2} g(\ln b), \epsilon b^{d} g(\ln b), 
   T b^d g(\ln b))\quad.
\label{eq:3.12a}
\eea
Notice that the scale dimension of $\Delta N$ is minus that of $t$, modulo the
logarithmic corrections to the latter. This leads to a resonance in the
RG flow equations for $\Delta N$, which in turn leads to an additional
logarithmic dependence of $\Delta N$ on $t$, see Appendix \ref{app:D}.
Anticipating that logarithm, we generalize Eq.\ (\ref{eq:3.12a}) to
\bea
\Delta N(t,\epsilon,T)&=&{\rm const.}\times t\,g(\ln b)\ln b
\nonumber\\
&&\hskip -32pt + b^{-(d-2)}{\tilde f}_{N}(t b^{d-2}g(\ln b), 
    \epsilon b^d g(\ln b), T b^d g(\ln b)) \quad.
\nonumber\\
\label{eq:3.12b}
\eea
\eml%
This implies
\bml
\label{eqs:3.13}
\bea
N(0,\epsilon,0)&=&N_{F}\,\left[1 + c_{N}\left(\frac{\epsilon}{\epsilon_{F}}
   \,g\left(\frac{1}{d}\ln\frac{\epsilon_{\rm F}}{\epsilon}\right)
       \right)^{(d-2)/d}\right.
\nonumber\\
&&\hskip 40pt + O(\epsilon^{(d-2)/2}\biggr]\quad,
\label{eq:3.13a}
\eea
and
\be
N(t,0,0) = N_{\rm F}\,\left[1 + d_N\,t\,g\left(\frac{1}{d-2}\ln\frac{1}{t}
   \right) \ln\frac{1}{t} + \ldots \right]\ ,
\label{eq:3.13b}
\ee
\eml%
with constants $c_N$ and $d_N$. It should be pointed out, however, that
one should not take the $tg(\ln 1/t)\ln t$ behavior in Eq.\ (\ref{eq:3.13b})
too seriously. The reason is that the log-log-normal factor $g(\ln 1/t)$ may
have multiplicative simple-log corrections that the asymptotic solution
of the field theory is not sensitive to, see the remark after
Eq.\ (\ref{eq:3.6e}).

\subsubsection{Transport coefficients}
\label{subsubsec:III.B.3}

The scaling theory for the transport coefficients, $\sigma$, $D$, and $D_{s}$
can be presented in several different ways. Here we give two arguments for
the scaling part of $\sigma$. The first one starts with the fact that the
conductivity is a charge current correlation function whose scale dimension
with respect to the quantum magnetic fixed point is expected to be zero.
That is, $\sigma$ neither vanishes nor diverges at this quantum critical
point. However, $\sigma$ will depend on the critical dynamics, since the
paramagnon propagator enters the calculation of $\sigma $ in perturbation
theory, see Appendix \ref{app:B}. 
The critical correction to the bare or background conductivity
further depends linearly on the leading irrelevant operator, which we denote by
$u$. The latter is related to diffusive electron dynamics, and one therefore
expects the scale of $u$ to be the same as in disordered Fermi liquid theory,
namely, $[u]=-(d-2)$.\cite{us_fermions} Scaling arguments then suggest a 
generalized homogeneity law
\bml
\label{eqs:3.14}
\bea
\sigma(t,T,\Omega)&=&f_{\sigma}(t b^{d-2} g(\ln b), T b^{d} g(\ln b), 
                       \Omega b^d g(\ln b),
\nonumber\\
&&\hskip 80pt u b^{-(d-2)})
\nonumber\\
&&\hskip -20pt ={\rm const.} + b^{-(d-2)}\,{\tilde f}_{\sigma}(t b^{d-2} 
    g(\ln b), T b^{d} g(\ln b),
\nonumber\\
&&\hskip 70pt \Omega b^d g(\ln b))\quad.
\label{eq:3.14a}
\eea
Again, there is a resonance condition that leads to a simple logarithm
in the $t$-dependence of $\sigma$. This is due to the
scale dimension of $u$ being minus that of $t$, modulo logarithmic
corrections. As in the case of the density of states, we therefore
generalize Eq.\ (\ref{eq:3.14a}) to
\bea
\sigma (t,T,\Omega)&=& {\rm const.} + {\rm const.}\times t\,g(\ln b)\ln b
\nonumber\\
&&\hskip -40pt + b^{-(d-2)}\,{\tilde f}_{\sigma }(t b^{d-2} g(\ln b), 
   T b^d g(\ln b), \Omega b^d g(\ln b))\quad.
\nonumber\\
\label{eq:3.14b}
\eea
\eml%
This scaling relation yields
\bml
\label{eqs:3.15}
\bea
\sigma(0,T,0)&=&\sigma_{0}\left[1 + c_{\sigma}\left(\frac{T}{T_{F}}\,
g\left(\frac{1}{d}\ln\frac{\epsilon_{F}}{T}\right)\right)^{(d-2)/d}\right.
\nonumber\\
&&\hskip 30pt +O(T^{(d-2)/2}) \biggr] \quad,
\label{eq:3.15a}
\eea
and
\be
\sigma(t,0,0) = \sigma_{0}\left[1 + d_{\sigma}t\,
   g\left(\frac{1}{d-2}\ln\frac{1}{t} \right) \ln\frac{1}{t} + O(t)\right]
   \quad,
\label{eq:3.15b}
\ee
\eml%
with $c_{\sigma}$ and $d_{\sigma}$ constants. The same caveat as given
after Eq.\ (\ref{eq:3.13b}) applies.

An alternative argument that gives Eqs.\ (\ref{eqs:3.15}) is to assume that 
$\sigma$ consists of a background part that does not scale, and a singular 
part, $\delta\sigma$, that does. In fundamental units, 
$[\delta\sigma] = -(d-2)$. This suggests the scaling form
\bml
\label{eqs:3.16}
\bea
\delta\sigma(t,T,\Omega)&=&b^{-(d-2)}
\nonumber\\
&&\hskip -42pt \times
   {\tilde f}_{\sigma}(t b^{d-2} g(\ln b), T b^{d} g(\ln b), 
    \Omega b^d g(\ln b))\quad.
\label{eq:3.16a}
\eea
Here the dependence on $u$ is already implicitly taken into account, so
we dropped the explicit dependence. Taking into account the resonance between 
the scale dimension of $\delta\sigma$ and $t$, see Appendix \ref{app:D}, gives
\bea
\delta\sigma(t,T,\Omega)&=&{\rm const.}\times t\,g(\ln b)\ln b
\nonumber\\
&&\hskip -42pt +b^{-(d-2)}\,{\tilde f}_{\sigma}(t b^{d-2} g(\ln b), 
  T b^{d} g(\ln b), \Omega b^d g(\ln b)) \quad.
\nonumber\\
\label{eq:3.16b}
\eea
\eml%
This also yields Eqs.\ (\ref{eqs:3.15}).

The diffusion coefficients $D$ and $D_{s}$ have dimensions of length squared
divided by time. Since there are two time scales, there 
are two possible scale dimensions for the diffusion
coefficients. $D_{s}$ is the diffusion coefficient for the
order parameter fluctuations, so one expects the critical time scale
to apply, while $D$ is the quasi-particle diffusion coefficient, so the 
diffusive time scale is appropriate. This leads to homogeneity laws
\bml
\label{eqs:3.17}
\bea
D_{s}(t,T,\Omega)&=&\left[b^{-(d-2)}/g(\ln b)\right]\,
\nonumber\\
&&\hskip -47pt \times f_{D_{s}}(t b^{d-2}g(\ln b), 
   T b^d g(\ln b), \Omega b^d g(\ln b))\quad,
\label{eq:3.17a}
\eea
\bea
D(t,T,\Omega)&=&[g(\ln b)]^{-1}
\nonumber\\
&&\hskip -44pt \times f_{D}(t b^{d-2} g(\ln b), T b^d f(\ln b),
   \Omega b^d g(\ln b))\quad.
\label{eq:3.17b}
\eea
\eml%
These two results are consistent with the fact that the conductivity is
noncritical to leading order, and scales like
$\sigma \sim D_{s}\chi_{s} \sim DH$.
Indeed, the above results for $\sigma$, $D_{s}$, and $D$ can 
be used to obtain scaling results for $\chi_{s}$ and $\gamma_C\propto H$, 
which justify our free energy considerations in Sec.\ \ref{subsubsec:III.B.2}
above.

\subsubsection{Relaxation rates}
\label{subsubsec:III.B.4}

The phase breaking rate $\tau_{\rm ph}^{-1}$ and the quasiparticle decay
rate $\tau_{\rm QP}^{-1}$ are given by Eqs.\ (\ref{eqs:3.4}).
By comparing Eq.\ (\ref{eq:3.4b}) and Eq.\ (\ref{eq:3.3a}), we see that
$\tau_{\rm ph}^{-1}$ scales like a wavenumber squared (recall that $G$ is
not singularly renormalized and hence does not scale), and therefore 
has a scale dimension
$[\tau_{\rm ph}^{-1}] = 2$ with no logarithmic corrections. This observation
leads to the homogeneity law
\bml
\label{eqs:3.18}
\bea
\tau_{\rm ph}^{-1}(t,\epsilon,T)&=&b^{-2}
\nonumber\\
&&\hskip -40pt\times f_{\rm ph}(t b^{d-2}g(\ln b),\epsilon b^d g(\ln b), 
   T b^d g(\ln b),u b^{-(d-2)})\ .
\nonumber\\
\label{eq:3.18a}
\eea
The leading irrelevant variable $u$ represents interaction effects that
are necessary for any dephasing. The rate is therefore linear in $u$,
and we can write
\bea
\tau_{\rm ph}^{-1}(t,\epsilon,T)&=&b^{-d}
\nonumber\\
&&\hskip -40pt\times {\tilde f}_{\rm ph}(t b^{d-2}g(\ln b),
   \epsilon b^d g(\ln b), T b^d g(\ln b))\ .
\label{eq:3.18b}
\eea
At criticality, we find
\be
\tau_{\rm ph}^{-1}(0,\epsilon,0) = c_{\rm ph}\,\epsilon\,
   g\left(\frac{1}{d}\ln\frac{\epsilon_{\rm F}}{\epsilon}\right)
   + O(\epsilon^{d/2}) \quad,
\label{eq:3.18c}
\ee
\eml%
with $c_{\rm ph}$ a constant.

The quasi-particle relaxation rate is given by the ratio $H''/H'$, see
Eq.\ (\ref{eq:3.4a}).
The scaling properties of $\tau_{\rm QP}^{-1}$ thus follow from those
of $H$, or $\gamma_C$, Eq.\ (\ref{eq:3.9b}). Explicitly we find
\be
\tau_{\rm QP}^{-1}(t=0,\epsilon,T=0) = c_{\rm QP}\,\epsilon \ln \left[
  \ln\frac{\epsilon_{\rm F}}{\epsilon}\right]/\ln \frac{\epsilon_{\rm F}}
   {\epsilon} + \ldots\quad,
\label{eq:3.19}
\ee
with $c_{\rm QP}$ a constant.
For $\epsilon\ll\epsilon_{\rm F}$, $t\neq 0$ we have asymptotically
\be
\tau_{\rm ph}^{-1}(t,\epsilon)\propto\tau_{\rm QP}^{-1}(t,\epsilon)\propto
   (\epsilon/t)^{d/2}\quad.
\label{eq:3.20}
\ee
To obtain Eq. (\ref{eq:3.20}) we have used the well-known fact 
that at the disordered
Fermi-liquid fixed point the relaxation rates are proportional to
$\epsilon^{d/2}$. In terms of our scaling arguments this result is
rederived in Appendix \ref{app:E}.

\section{Discussion, and experimental relevance}
\label{sec:IV}

This paper completes our discussion of various aspects of the ferromagnetic
phase transition in low-temperature disordered itinerant electron systems.
We have given the exact solution for the magnetic critical behavior near the
quantum phase transition from a paramagnetic metal to a ferromagnetic metal.
We have also determined the critical behavior of a number of other relevant
physical variables near the transition, in particular the electrical
conductivity and the tunneling density of states. In addition, 
we have established
several connections between previously formulated theories. We conclude
with a discussion of general aspects of our results, and their experimental
relevance, starting with the former.

\subsection{General discussion}
\label{subsec:IV.A}

One important aspect of the present analysis is the establishment of 
connections 
between various theoretical formulations of the quantum ferromagnetic
transition problem. First, we related our previous nonlocal order parameter 
field theory\cite{us_dirty} to the local coupled field theory involving 
both the fermion density fluctuations and the magnetic order parameter 
fluctuations that was formulated in I. The two theories yield the same critical 
behavior apart from logarithmic corrections to power-law scaling that were
missed in Ref.\ \onlinecite{us_dirty}, and are correctly given by the coupled 
local theory. Second, we have unambiguously related earlier work on the 
disordered interacting electron problem, involving runaway renormalization 
group flow, to the quantum ferromagnetic phase 
transition.\cite{runaway_footnote} This connection was actually made earlier 
in Ref.\ \onlinecite{us_IFS}. However, as explained in the Introduction, 
our argument then was not complete. Indeed, even though 
Ref.\ \onlinecite{us_IFS} correctly obtained the critical behavior,
it failed to identify the nature of the phase transition as the ferromagnetic 
one. In hindsight, this is surprising, given that scaling theory was used to 
correctly identify the order parameter exponent $\beta = 2\nu$. The problem 
was that Ref.\ \onlinecite{us_IFS} was formulated solely in terms of fermionic
number and spin density fluctuations at the Fermi surface, so that the behavior
of quantities that involve electrons far from the Fermi surface, such as the
magnetization, was not obvious. Indeed, we argued that although the above 
exponent equality  was formally valid, the magnetization was actually
zero in the ordered phase since the scaling function had a zero 
prefactor.\cite{R_footnote} This argument was incorrect because it failed to 
take into account that electrons away from the Fermi surface are also ordering.
This, in turn, leads to a nonzero scaling function.

The difficulties interpreting the theory put forward in 
Ref.\ \onlinecite{us_IFS}
notwithstanding, it is very remarkable that this nonlinear sigma model
formulation of the problem yielded the correct result, since it was not
geared at all towards describing ferromagnetism. The focus on degrees of
freedom near the Fermi surface mentioned above is one reason, and another
one is the fact that the sigma model is derived by expanding about the
paramagnetic metal fixed point, so it is not obvious why it is capable of
describing a critical fixed point. This is actually a general question about
sigma models,\cite{ZJ} and the answer is only incompletely known. Indeed,
periodically even the capability of the $O(N)$ nonlinear sigma model to
qualitatively correctly describe the Heisenberg transition in $d=3$ has been
questioned.\cite{Heisenberg_FP}

The connection between the runaway renormalization group flow encountered
in low orders of a loop expansion and
ferromagnetism is particularly interesting in two-dimensional systems
because of recent  experiments that show either metallic or metallic-like
behavior in Si MOSFETs and other materials,\cite{AKS} and even more recent 
ones that show that this behavior happens near a quantum phase transition to a
ferromagnetic state.\cite{2d_fm} The connection between ferromagnetism and
two-dimensional metallic-like behavior is not obvious, but the observation
of a ferromagnetic phase in $d=2$ is consistent with our identification of the
runaway flow behavior with ferromagnetism. The same connection was more
recently made by others.\cite{Chamon} It is also interesting
to note that a nearby ferromagnetic phase in disordered systems is favorable 
to an exotic type of even-parity, triplet 
superconductivity.\cite{us_triplet_sc} After the experimental observation
of $2$-$d$ metallic behavior, this was proposed as a possible 
explantion.\cite{us_MOSFET} Proposals of superconducting or otherwise exotic
phases as the explanation for the observations are bolstered by the conclusion
that conventional metallic behavior in a two-dimensional ferromagnetic system
is unlikely to occur.\cite{us_fm_mit_I} Even if the observed metallic or
pseudo-metallic phase is unrelated to superconductivity, the presence of a
ferromagnetic phase makes the existence of a triplet superconducting phase
nearby more likely.

\subsection{Experimental consequences}
\label{subsec:IV.B}

Most of the results of the present paper can be directly checked by
experiments, at least in principle. For instance, the pressure tuned
ferromagnetic transitions observed at very low temperatures in 
MnSi\cite{Lonzarich_I} and UGe$_2$\cite{Lonzarich_II} provide examples
of systems where the quantum critical point is directly accessible.
These experiments were done on very clean samples, where the ferromagnetic
transition at low temperatures is of first order, in agreement with
theoretical predictions.\cite{us_1st_order} However, upon introducing
quenched disorder one expects the transition to become of second 
order,\cite{us_1st_order} and the current theory to apply.

The critical behavior predicted for the thermodynamic quantities is
markedly different from the mean-field exponents predicted by
Hertz's theory.\cite{Hertz} For instance, the predicted value of the
magnetization exponent in $d=3$, $\beta=2$ with logarithmic corrections, is
very different from both the mean-field value $\beta_{\rm MF}=0.5$
and the $3$-$d$ classical Heisenberg value $\beta_{\rm H} \approx 0.37$.
One important remark in this context is that the logarithmic corrections
will, over any realistically achievable range of $t$-values, mimic a
power, so that the observed value of $\beta$ should be expected to be
smaller than $2$. Similarly, the correlation length exponent $\nu$
will be equal to $1$ with logarithmic corrections in $d=3$. However,
these exponents may be hard to measure directly, especially near a
quantum phase transition that must be triggered by a non-thermal control
parameter that is more difficult to accurately vary than the temperature. It
is therefore important that the values of the critical exponents are
also reflected in the behavior of the tunneling density of states,
and the electrical conductivity, across the transition. Even though
these observables do not show any leading critical behavior, the
leading corrections expressed in Eqs.\ (\ref{eqs:3.13}) and (\ref{eqs:3.15})
reflect the values of $\nu$ and $z$, and they should be easier to measure
than the critical behavior of, say, the magnetization. For instance,
our prediction for the tunneling density of states in $d=3$ is as follows.
Far from the transition, it will show the well-known square-root
anomaly as a function of the bias voltage that is characteristic of
disordered metals.\cite{AA} Near the transition, the voltage region
that shows the square-root behavior will shrink, and outside of it
a region of cube-root behavior will appear, until at criticality
the behavior is given by the $\epsilon^{1/3}$ behavior shown in
Eq.\ (\ref{eq:3.13a}). The same discussion applies to the conductivity
as a function of temperature, see Eq. (\ref{eq:3.15a}). Again, the
logarithmic corrections to scaling will manifest themselves in a
real experiment as an effective power smaller than $1/3$.

In this context we also come back to the relative values of the
coefficients $a_2$ and $a_{d-2}$ in Eq.\ (\ref{eq:2.3a}).
The leading critical behavior is due to the $\vert{\bf k}\vert^{d-2}$
term whose coefficient is $a_{d-2}$, as can be seen, for instance,
from Eq.\ (\ref{eq:B.2}) in conjunction with Eq.\ (\ref{eq:2.6c}).
Since $a_{d-2} = O(1/k_{\rm F}\ell)$, see Eq.\ (\ref{eq:2.3a}),
this implies that the leading effects will be manifest only for
sufficiently strong disorder, or, if we scale the wavenumber with
the correlation length $\xi$, for sufficiently large $\xi$ at fixed
mean-free path $\ell$. Since $a_d = O(1)$, the nonanalytic term will
dominate for $\xi\agt\ell$ or, using $\nu=1$, for $t\alt 1/k_{\rm F}\ell$.
Typical values of the disorder results in mean-free paths 
$\ell\approx 10/k_{\rm F}$. For such a value, the our leading results
will apply everywhere in the critical region. For less disordered
samples, their region of validity will be correspondingly narrower.

\subsection{Conclusion}
\label{subsec:IV.C}

In conclusion, we now have a complete theory for the quantum critical 
behavior of disordered itinerant ferromagnets in $d>2$, including the exact
values of the critical exponents, the leading logarithmic corrections 
to power-law scaling, and the relations between various theoretical
approaches to the problem. Specific predictions for the behavior of all
important observables allow for a direct experimental test of this theory.
However, in $d=2$ there remains a puzzling discrepancy between existing
theory and observations. The latest experimental evidence is for a 
transition, with increasing electron density, from a paramagnetic insulator 
to a ferromagnetic metal,\cite{2d_fm} while there is no theory that can
account for a metallic state, ferromagnetic or otherwise, in $d=2$. In
particular, it has recently been shown that ferromagnetic fluctuations
in $d=2$ do not produce a metallic state within a perturbative RG 
treatment,\cite{us_fm_mit_I} ruling out a possible mechanism for a
metal-insulator transition in $d=2$. This state of affairs has recently
been reviewed in Ref.\ \onlinecite{AKS}.

\acknowledgments
Part of this work was performed at the Aspen Center for Physics. We thank
the Center for hospitality, and E. Abrahams for helpful discussions.
This work was supported by the NSF under grant numbers DMR-98-70597 and
DMR-99-75259.

\appendix
\section{Additional 3-point and 4-point vertices}
\label{app:A}

\begin{figure}[t,h]
\centerline{\psfig{figure=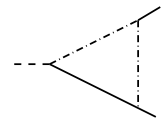,width=20mm}\vspace*{5mm}}
\caption{A one-loop diagram that generates a new three-point vertex. Parts
 of the diagrams shown in Fig.\ \ref{fig:2} also contribute to this vertex.}
\label{fig:A1}
\end{figure}

Starting at one-loop order, the RG generates vertices that are not in
the effective action. An example is the 3-point vertex shown in
Fig.\ \ref{fig:A1}.
In contrast to the vertex $c_2$, all external legs in this diagram carry 
the same replica index. To see the physical meaning of this term, we
integrate out the $b$-field to arrive at an effective 4-point $q$-vertex
that is completely diagonal in replica space.
This corresponds to a four-body interaction term which in imaginary time
space must have the form
\bml
\label{eqs:A1}
\be
\int_0^{\beta} d\tau\ \left(n_{\rm s}(\tau)\right)^4\quad,
\label{eq:A1a}
\ee
with $n_{\rm s}(\tau)$ an electron spin density in imaginary time 
representation. Performing a Fourier transform, and making use of the 
isomorphism between density operators and
$q$-matrices that has been explained in I (see also 
Ref.\ \onlinecite{us_fermions}),
this corresponds, in schematic notation, to
\be
T^3\int d{\bf x} \sum_{\alpha}\left(Q^{\alpha\alpha}({\bf x})\right)^4\quad.
\label{eq:A1b}
\ee
\eml%
Here $Q({\bf x})$ is the matrix field from I Eq.\ (2.8), and we have suppressed
frequency labels and frequency sums for clarity. A Hubbard-Stratonovich
transformation to reintroduce the $b$-field then leads to a $b q^2$ vertex
of the structure
\be
{\tilde c}_2\,T^{3/2}\int d{\bf x}\sum_{\alpha} b^{\alpha\alpha}({\bf x})\,
   q^{\alpha\alpha}({\bf x})\,q^{\alpha\alpha}({\bf x})\quad,
\label{eq:A2}
\ee
with ${\tilde c}_2$ a coupling constant. This vertex thus carries a higher
power of the temperature than the one with coupling constant $c_2$. 
By using Eqs.\ (\ref{eqs:2.11})
to estimate the behavior of the diagram shown in Fig.\ \ref{fig:A1} we
find that ${\tilde c}_2$ diverges for $d<4$ in the long-wavelength and small
frequency limit, ${\tilde c}_2 = {\tilde{\tilde c}}_2\,\Lambda^{d-4}$, with
$\Lambda$ the infrared momentum cutoff from Sec.\ \ref{subsubsec:II.B.2}.
The scale dimension of ${\tilde{\tilde c}}_2$ is therefore smaller than that of
$c_2$,
\be
[{\tilde{\tilde c}}_2] \leq [c_2] - (d-2)\quad.
\label{eq:A3}
\ee
This term is therefore irrelevant for the critical behavior.

\begin{figure}[t,h]
\centerline{\psfig{figure=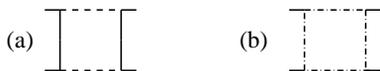,width=50mm}\vspace*{5mm}}
\caption{One-loop diagrams that generate (a) a two-body and (b) a four-body
 interaction.}
\label{fig:A2}
\end{figure}

Similarly, the RG generates four-point vertices that are not in the effective
action. For instance, diagram (a) in Fig.\ \ref{fig:A2} regenerates the 
two-body interaction that was shown in I to be irrelevant and therefore 
dropped. Diagram (b) is a four-body interaction of the same type as discussed
above in connection with Fig.\ \ref{fig:A1}. All of these terms, and similar
ones not shown in Fig.\ \ref{fig:A2}, are thus
RG irrelevant and can be safely neglected.

\section{Perturbative results for the density of states and the conductivity}
\label{app:B}

In this Appendix we show that the scaling results
obtained in Sec.\ \ref{sec:III} are consistent with perturbation theory
for $N$ and $\sigma$.

From Eq. (\ref{eq:3.1}) we find the one-loop result for $\Delta N$ as,
\bea
\Delta N(\epsilon_{\rm F} + \epsilon)&=&- \frac{N_{\rm F}}{2}\sum_{\beta}\sum_m
   \sum_{i,r}\langle {^i_rq}_{nm}^{\alpha\beta}({\bf x})\,
\nonumber\\
&&\hskip 40pt \times
         {^i_rq}_{nm}^{\alpha\beta}({\bf x})
          \rangle_{i\omega_{n}\rightarrow \epsilon +i0}\quad.
\label{eq:B.1}
\eea
Using Eqs.\ (\ref{eq:2.7}), (\ref{eqs:2.8}), this yields
\bea
\Delta N(\epsilon_{\rm F} + \epsilon)&=&\frac{3\pi}{4}\,G^2 N_{\rm F}
   K_t\frac{1}{V}
   \sum_{\rm p} T\sum_{m<0} \left({\cal D}_{n-m}({\bf p})\right)^2
\nonumber\\
&&\hskip 40pt \times {\cal M}_{n-m}({\bf p})
      \Bigl\vert_{i\omega_{n}\rightarrow \epsilon +i0}\quad.
\label{eq:B.2}
\eea
In Sec.\ \ref{sec:II} we have seen that self-consistent one-loop theory
for the propagators $\cal D$ and $\cal M$ is exact. Using the resulting
dressed propagators in Eq. (\ref{eq:B.2}) and doing the integrals, one
obtains Eqs. (\ref{eqs:3.13}).

Similarly, if we use the perturbative result for the conductivity given by
Eqs. (3.6) of I and the exact propagators derived in 
Sec.\ \ref{sec:II} of the present paper, then Eqs. (\ref{eqs:3.15}) for 
$\sigma$ are obtained.

These two results buttress the scaling arguments given in Sec. \ref{sec:III}.
Note that these perturbative calculations are equivalent to taking the
standard Altshuler-Aronov perturbative results,\cite{AA} and using 
propagators that are appropriate near the magnetic quantum phase transition.

The classical limit of Eq.\ (\ref{eq:B.2}) 
for the density of states, or the equivalent
result for the conductivity $\sigma $, can be related to the established
result for the critical behavior of the conductivity at a Heisenberg
critical point.\cite{FisherLanger} To see this, note that
at finite temperatures the diffusion propagator, ${\cal D}$ in 
Eq.\ (\ref{eq:B.2}), has
a mass due to inelastic scattering processes. For the leading critical
behavior, ${\cal D}$ can therefore be replaced by a constant. Also, in the
classical limit, the frequency sum in Eq.\ (\ref{eq:B.2}) turns into an
integral over all frequencies, since in
this limit the Boltzmann weight that restricts the frequency sum is
absent. The net result is that this contribution to $\Delta N$ or 
$\Delta\sigma$ is proportional to a magnetization-magnetization correlation 
function that is local in space and time,
\be
\Delta N\propto\Delta\sigma\propto\langle{\bf M}^2({\bf x},\tau)\rangle\quad.
\label{eq:B.3}
\ee
The correlation function on the right hand side of Eq.\ (\ref{eq:B.3}) 
is essentially the magnetic energy density,
and hence scales as $t^{1-\alpha}$, with $t$ the distance in
{\em temperature space} to the classical phase transition, and $\alpha$ the
usual specific heat critical exponent. This gives the result of
Ref.\ \onlinecite{FisherLanger},
\be
\Delta N\sim \Delta\sigma \sim t^{1-\alpha}\quad. 
\label{eq:B.4}
\ee

\section{Equation of state}
\label{app:C}

In this Appendix we show how to generalize the self-consistent equations
given in Sec.\ \ref{sec:II} for correlation functions in the paramagnetic
phase to the ferromagnetic phase. In this way we will derive and
validate the scaling argument given in Sec.\ \ref{sec:III} 
for the magnetization $m$ as a function of $t$.

To simplify the discussion we first ignore the logarithmic corrections to
scaling. Then, in the paramagnetic phase, the diffusion coefficient $D$
is simply a number, and the equation for $D_{s}$ is given by
Eq.\ (\ref{eq:2.14a}) with $D(\omega)$ replaced by that number.
The propagators in the ordered phase have been derived in
Ref.\ \onlinecite{us_fm_mit_I}. 
Assuming that the magnetization is in the $z$-direction, we first
need to specify whether longitudinal or transverse spin density correlations
will be considered. In the ferromagnetic phase, the transverse spin-density 
fluctuations become propagating Goldstone modes or spin wave excitations,
while the longitudinal ones remain diffusive. For the longitudinal spin-density
mode, the left hand side of Eq.\ (\ref{eq:2.14a}) therefore still describes
a diffusion coefficient. It is easily shown that in this case the
diffusion poles on the right hand side of this equation are cut off by the
magnetization. The frequency corresponding to this cutoff is a ``cyclotron''
frequency that scales like $\omega_{c}\sim \vert m\vert$. The explicit
generalization of these propagators to the ferromagnetic phase is given
by Eqs. (3.10) of Ref.\ \onlinecite{us_fm_mit_I}. 
Using these results, the generalization of Eq.\ (\ref{eq:2.14a}) 
to the ferromagnetic phase is, at zero external frequency and wavenumber,
\be
D_s = D_s^0 + \frac{iG}{2V}\sum_{\bf p}\int_0^{\infty}d\omega\,
   \frac{1}{({\bf p}^2 - i\omega/D + c\vert m\vert)^2}\quad,
\label{eq:C.1}
\ee
with $c$ a constant. Carrying out the integrals yields, for $2<d<4$,
\be
D_s = -c_1\vert t\vert + c_2\,m^{(d-2)/2}\quad.
\label{eq:C.2}
\ee
Here $\vert t\vert = -t > 0$ is the distance from criticality in the
ferromagnetic phase, and $c_1$ and $c_2$ are constants.
Equation (\ref{eq:C.2}) implies that $m$ scales as
\be
m\sim \vert t\vert^{2/(d-2)}\quad.
\label{eq:C.3}
\ee
Equation\ (\ref{eq:C.3}) is consistent with the scaling result for the
critical exponent $\beta$, Eq. (\ref{eq:3.11a}), apart from
logarithmic terms.

The logarithmic corrections to the exponent $\beta$ can be understood
as follows. As noted in Sec.\ \ref{subsubsec:III.B.1}, the correlation 
length exponent $\nu$ has leading logarithmic corrections, while the critical
exponent $\gamma$ does not.\cite{us_IFS} Within the integral equation approach,
and in the paramagnetic phase, this manifests itself in the following structure
of the renormalized paramagnon propagator (see Eqs. (\ref{eq:3.5}) and
(\ref{eq:2.17b})),
\be
{\cal M}({\bf k},i\Omega_n) = \frac{1}{t + \vert{\bf k}\vert^{d-2}/
   g(\ln(k_{\rm F}/\vert{\bf k}\vert)) + \vert\Omega_n\vert/{\bf k}^2}\quad.
\label{eq:C.4}
\ee
That is, the term proportional to $\vert{\bf k}\vert^{d-2}$ carries leading
logarithmic corrections, while the term $t$ does not. In Eq.\ (\ref{eq:C.4})
we have left out all constants for clarity. If one scales the wavevector
with the correlation length, then this structure produces the logarithmic
corrections to the exponent $\nu$. In the ferromagnetic phase, the
$\vert{\bf k}\vert^{d-2}$ nonanalyticity is cut off by a magnetic length
or cyclotron radius, $\ell_m \propto 1/m^{1/2}$. This means that the
$\vert{\bf k}\vert^{d-2}$ gets replaced by $m^{(d-2)/2}$, see 
Eq.\ (\ref{eq:C.2}). The net result is that, again, $t$ has no leading
logarithmic correction, while $m^{(d-2)/2}$ does have one. Scaling $m$
with the appropriate power of $t$ then yields Eq.\ (\ref{eq:3.11a}) for
$\beta$.

\section{Logarithmic corrections to scaling}
\label{app:D}

Wegner\cite{Wegner} has given a classification of logarithmic corrections
to scaling. The first class consists of simple logarithms that arise due
to resonance conditions between scale dimensions. In the present context,
such a resonance occurs between the scale dimensions of the leading correction
to the single-particle density of states and the relevant variable $t$. For
clarity, let us neglect the more complicated logarithms that are embodied 
in the function $g(\ln b)$ for the time being. The one-loop flow equations 
for these two quantities then are
\bml
\label{eqs:D.1}
\bea
\frac{d\,\Delta N}{d\ln b}&=&(d-2)\Delta N + {\rm const.}\times t\quad,
\label{eq:D.1a}\\
\frac{dt}{d\ln b}&=&(d-2)t\quad.
\label{eq:D.1b}
\eea
\eml%
The general solution of the homogeneous equation for $\Delta N$ is
\be
(\Delta N)_{\rm hom}(b) = (\Delta N)(b=1)\,b^{d-2}\quad.
\label{eq:D.2}
\ee
This has the same $b$-dependence as the inhomogeneity, $t(b)$. 
Consequently, the solution of the inhomogeneous equation is
\be
(\Delta N)(b) = \left[(\Delta N)(b=1) + {\rm const.}\times t(b=1)\ln b\right]\,
   b^{d-2}\ .
\label{eq:D.3}
\ee
The resulting logarithm has been taken into account in Eq.\ (\ref{eq:3.12b}).
(Notice that the physical quantity is $\Delta N (b=1)$.)
The same mechanism is at work for the conductivity, and this is reflected
in Eq.\ (\ref{eq:3.14b}).

Wegner's second mechanism is due to marginal operators, and it can lead
to arbitrary functions of logarithms. In our case, $c_2$ acts as an
effectively marginal operator, as has been explained in I, and this leads
to the log-log-normal factors we denote by $g(\ln b)$.

\section{Relaxation rates in a Fermi liquid}
\label{app:E}

Here we illustrate how to obtain Schmid's result\cite{Schmid} 
for the relaxation rates
in a disordered Fermi liquid from the scaling theory developed in
Sec.\ \ref{sec:III}.

At a disordered Fermi liquid fixed point, the dynamical exponent is $z=2$,
reflecting the diffusive dynamics of the quasiparticles, and the leading
irrelevant variable, which we denote by $u$, has a scale dimension
$[u] = -(d-2)$.\cite{us_fermions} Since
$\tau_{\rm ph}^{-1}\sim\tau_{\rm QP}^{-1}$ both are dimensionally
frequencies or energies, they scale the same way and we have for
either rate a homogeneity law
\be
\tau^{-1}(\epsilon,T) = b^{-2}\,f_{\tau}(\epsilon b^2, T b^2, ub^{-(d-2)})
   \quad.
\label{eq:E.1}
\ee
The dependence on $u$ arises from the electron-electron interaction terms
that lead to $\tau^{-1}\neq 0$ in the first place, and therefore the scaling
function has the property $f_{\tau^{-1}}(1,0,x)\propto x$. The explicit
dependence on $u$ can therefore be eliminated by writing, instead of
Eq.\ (\ref{eq:E.1}),
\bml
\label{eqs:E.2}
\be
\tau^{-1}(\epsilon,T) = b^{-d}\,{\tilde f}_{\tau}(\epsilon b^2, T b^2)\quad.
\label{eq:E.2a}
\ee
In particular, we have\cite{Schmid}
\be
\tau^{-1}(\epsilon,0) = {\tilde f}_{\tau}(1,0)\,\epsilon^{d/2} \quad,
\label{eq:E.2b}
\ee
\eml%
which we used to derive Eq.\ (\ref{eq:3.20}).

\end{document}